\documentclass[letterpaper, 10 pt, conference]{ieeeconf}  

\IEEEoverridecommandlockouts                              
                                                          
\usepackage{graphicx}
\usepackage{amsmath}
\usepackage{cite}
\usepackage{tabulary}
\usepackage{amssymb}
\usepackage{subcaption}

\usepackage{multirow}
\usepackage{booktabs}

\title{\LARGE \bf
Grading the severity of hypoxic-ischemic encephalopathy in newborn EEG using a convolutional neural network}

\author{Sumit A. Raurale$^{1,2}$ \textit{Member, IEEE}, Geraldine B. Boylan$^{1,2}$, Gordon Lightbody$^{1,3}$ and \\ John M.
  O'Toole$^{1,2}$ \textit{Member, IEEE} 
  \thanks{This work is supported by the Wellcome Trust (209325/Z/17/Z) and Science Foundation
    Ireland (INFANT-12/RC/2272).  
    JMOT is supported by Science Foundation Ireland (15/SIRG/3580 and
    18/TIDA/6166).}
  \thanks{$^{1}$INFANT Research Centre, Ireland. 
    {\tt\small (sumit.raurale@ucc.ie)}}%
  \thanks{$^{2}$Department of Paediatrics and Child Health, University College Cork,
    Ireland.}%
  \thanks{$^{3}$Department of Electrical and Electronic Engineering, University College
    Cork, Ireland.}%
}

\begin{document}

\maketitle
\thispagestyle{empty}
\pagestyle{empty}

\begin{abstract}

Electroencephalography (EEG) is a valuable clinical tool for grading injury caused by lack of blood and oxygen to the brain during birth. This study presents a novel end-to-end architecture, using a deep convolutional neural network, that learns hierarchical representations within raw EEG data. The system classifies 4 grades of hypoxic-ischemic encephalopathy and is evaluated on a multi-channel EEG dataset of 63 hours from 54 newborns. The proposed method achieves a testing accuracy of 79.6\% with one-step voting and 81.5\% with two-step voting. These results show how a feature-free approach can be used to classify different grades of injury in newborn EEG with comparable accuracy to existing feature-based systems. Automated grading of newborn background EEG could help with the early identification of those infants in need of interventional therapies such as hypothermia.

\end{abstract}

\vspace{0.15cm}
\section{Introduction}
\label{introduction}
Electroencephalography (EEG) is a non-invasive tool that can continuously monitor brain function. For newborn infants, distinct EEG activity can provide information on sleep--wake cycling \cite{R3}, maturation \cite{R5}, potential neurodevelopmental outcome \cite{R4}, and hypoxic-ischemic encephalopathy (HIE) \cite{R10}. HIE is a type of injury caused by lack of oxygen (hypoxia) and impairment to blood supply (ischemia) to the brain during birth. In the clinical care of HIE infants, it is helpful to quantify the severity of HIE. The continuous, cotside recordings from EEG provide a rich source of information to grade the severity of the HIE \cite{R10}. However, continuous EEG recording and review is not practical in most neonatal intensive care units (NICUs) because of the lack of specialised expertise required to review neonatal EEG. Automated analysis of HIE grading could make continuous, cotside EEG monitoring possible for all infants in need without placing a significant burden on the existing clinical neurophysiology service. This would help ensure early and accurate stratification of infants and thus help improve health outcomes from these brain injuries.

EEG characteristics of HIE in term infants are numerous and varied, including, for example, absence of sleep--wake cycling, periods of low voltage activity, discontinuous activity with periods of suppression, and asymmetry between hemispheres \cite{R10}. 
Existing grading algorithms employ complex feature sets to extract information from the raw EEG \cite{R6,R12,R11}, before combining with machine learning methods. These extensive feature sets are needed to capture the wide range of signal characteristics associated with the different grades. Although we have recently developed a simple grading method that detects and quantifies inter-burst intervals in the EEG \cite{R13}, to generate a full grading system more quantitative EEG features are needed. Eschewing the feature-based approach, in this paper we apply a deep-learning approach which uses all information from the raw EEG signal without the need to design and select features.

Convolutional neural networks (CNN) have demonstrated state-of-the-art results in many medical image processing tasks. These CNNs were developed specifically for two-dimensional image data \cite{mimage}. However, there is growing evidence that CNNs can also be used to process one-dimensional (1D) medical time-series data such as EEG \cite{seizure,cnnsleep}. Thus, here we explore the use of a CNN model for HIE grading of full-term EEG.

\section{Methods}
\begin{figure*}[h]
	\centering
	\includegraphics[width=1.0\linewidth]{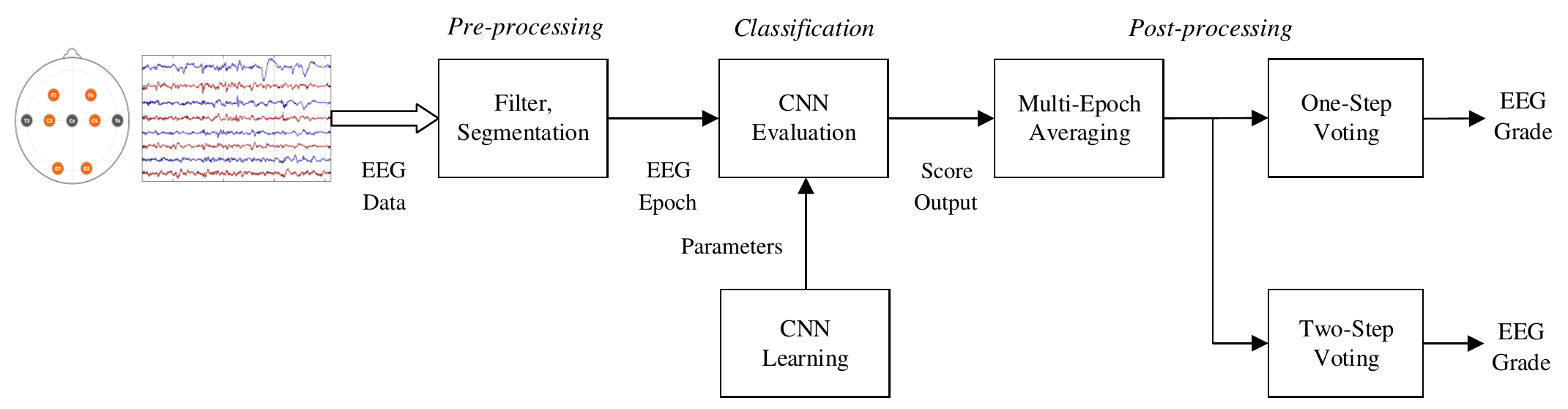}
	\caption{Proposed HIE grading system for term infant EEG.}
	\label{system}
\end{figure*}

We aim to develop an automated system using deep CNN to differentiate 4 grades of HIE using the EEG. This data-driven approach automatically extracts features and classifies the 4 grades of HIE from approximately 1 hour of multi-channel EEG. The system architecture for the proposed method is outlined in Fig. \ref{system}.


\subsection{EEG Data and Pre-processing}
The EEG was recorded from a NicoletOne EEG system from 54 term infants in the NICU of Cork University Maternity Hospital, Cork, Ireland. This study was approved by the Clinical Ethics Committee of the Cork Teaching Hospital with written and informed parental consent obtained before EEG recording. To monitor the evolution of the developing encephalopathy, EEG recordings were initiated within 6 hours of birth and continued for up to 72 hours. Nine active electrodes over the frontal, central, temporal, and occipital areas were used for EEG recording. Our analysis used an 8-channel bipolar montage derived from these electrodes as F4-C4, F3-C3, C4-O2, C3-O1, T4-C4, C3-T3, C4-Cz and Cz-C3. 

\begin{figure}[!h]
\vspace{0.15cm}
\centering
\begin{subfigure}[b]{0.238\textwidth}
   \includegraphics[width=1\linewidth]{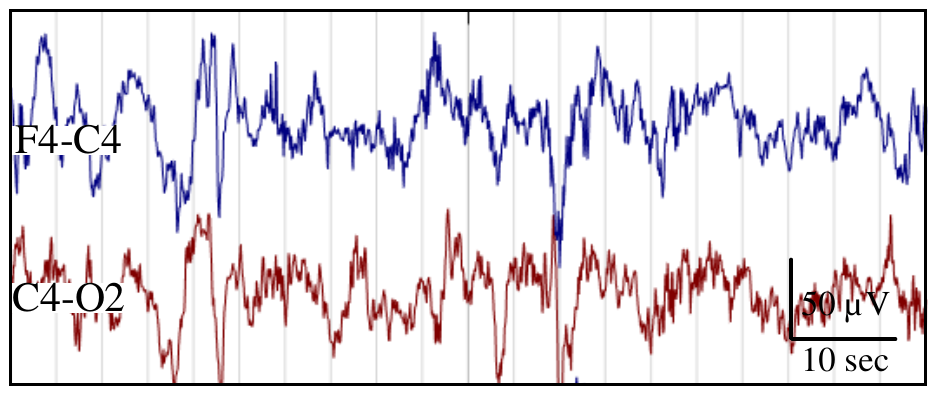}
   \caption{Normal/Mild abnormalities}
   \label{fig:Ng1} 
   \vspace{0.30cm}
\end{subfigure}
\begin{subfigure}[b]{0.238\textwidth}
   \includegraphics[width=1\linewidth]{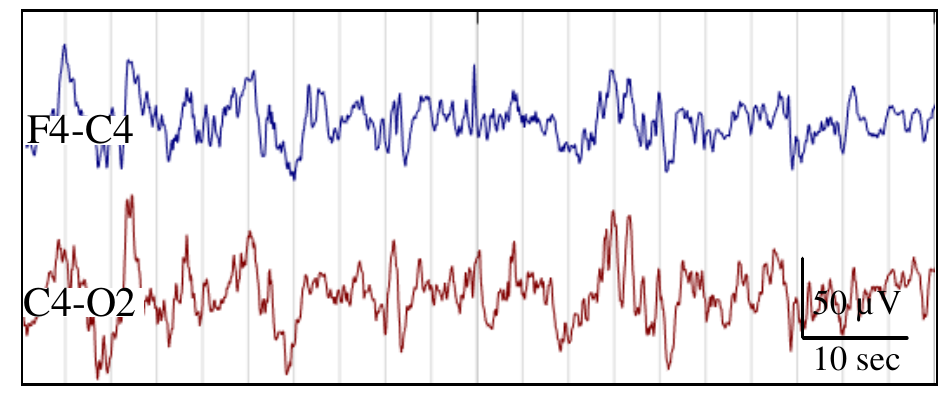}
   \caption{Moderate abnormalities}
   \label{fig:Ng2}
   \vspace{0.30cm}
\end{subfigure}
\begin{subfigure}[b]{0.238\textwidth}
   \includegraphics[width=1\linewidth]{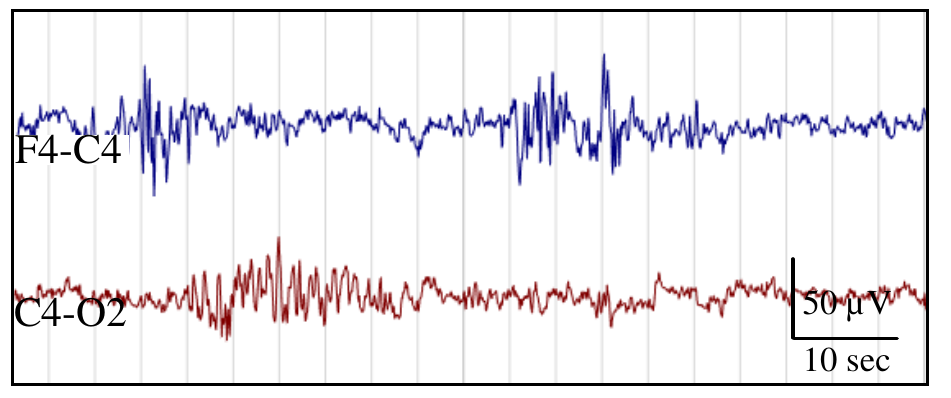}
   \caption{Major abnormalities}
   \label{fig:Ng3}
\end{subfigure}
\begin{subfigure}[b]{0.238\textwidth}
   \includegraphics[width=1\linewidth]{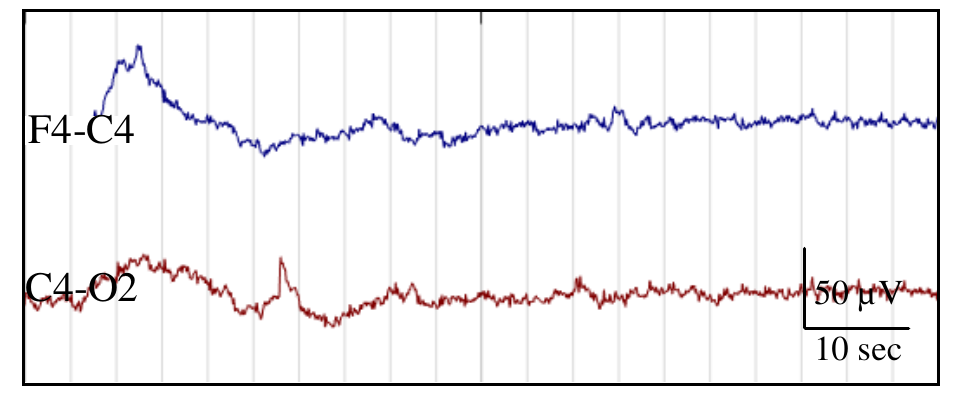}
   \caption{Severe abnormalities}
   \label{fig:Ng4}
\end{subfigure}
\caption{Examples of 2 channels of EEG for term infants for each of the 4 HIE grades.}
\label{fig_sim}
\vspace{-0.25cm}
\end{figure}

To avoid major artefacts, approximately 1-hour EEG epochs were pruned from the continuous EEG. Two EEG experts independently reviewed each epoch and graded according to the system defined by Murray \emph{et al.} \cite{murray}. In cases of disagreement, the experts jointly reviewed the EEG to reach consensus for the final grade. An EEG example with the 4 levels of HIE severity (grades) is shown in Fig. \ref{fig_sim}. This same dataset has been used by Raurale \emph{et al}. \cite{R13}, Ahmed \emph{et al}. \cite{R12}, and Stevenson \emph{et al}. \cite{R11}.

The only processing applied to the EEG was downsampling from 256 Hz to 64 Hz, after application of an anti-aliasing low-pass filter with a cut-off of 30Hz.

\subsection{CNN Architecture}
The proposed CNN architecture comprises of convolutional, rectifying, pooling, and normalisation layers; then followed by two fully-connected layers with a softmax layer.
The convolution layer is an implementation of a filter-bank, using finite-impulse response filters
$y_k = \bold{f}_k \ast \bold{x}$ for the $k$-th filter with $\bold{x}$ as the input vector and $\ast$ as the convolution operation.
The $k$-th vector $\bold{f}_k$, filter coefficients including a bias term, are learnt during training.

Convolution layers are followed by a rectified linear unit (ReLu) nonlinearity $\text{max}(0, y_k)$, which substitutes negative values with zero while keeping positive values unchanged. Next, a pooling layer is included to reduce the amount of parameters to train and to limit overfitting after each successive convolutional--ReLu layers. A pooling layer aggregates by downsampling the volume of the previous layer. Two pooling operators are used for downsampling: one takes the maximum value and the other averages over a prescribed window. A batch-normalization layer is also used to normalise the extracted feature maps after the first pooling layer.

The classification is performed by 2 fully connected layers, similar to the hidden layers in a classic multi-layer perceptron (MLP). Finally, the softmax layer is used to exponentially normalise the network outputs and represent them as probabilities corresponding to the target classes. The detailed layer structure for the proposed CNN architecture is illustrated in Table \ref{cnnarch}.

\begin{table}[h]
\renewcommand{\arraystretch}{1.0}
\centering
\caption{Layer configuration for the proposed CNN architecture.}
\label{cnnarch}
\begin{tabular}{c|l l l l}
\toprule
\multicolumn{1}{c}{} & \multicolumn{2}{c}{\textbf{Layer Info}} & \textbf{Activation} & \textbf{Parameters}\\
\midrule
\parbox[t]{2mm}{\multirow{15}{*}{\rotatebox[origin=c]{90}{Feature Extraction}}}  & -- & 1D Input & $19200\times1$ & -- \\
 & 1 & Conv (64,1) $\times$~10 & $\medmuskip=0mu 19200\times1\times10$ & $w\!:\!640,b\!:\!10$\\
 & 2 & ReLu & $\medmuskip=0mu 19200\times1\times10$~ & --\\
 & 3 & MPool (4,1), $s\!:$ 4 & $\medmuskip=0mu 4800\times1\times10$ & --\\
 & 4 & BNorm & $\medmuskip=0mu 4800\times1\times10$ & $o\!:\!10,c\!:\!10$\\
 & 5 & Conv (32,1) $\times$~20 & $\medmuskip=0mu 4800\times1\times20$ & $w\!:\!6400,b\!:\!20$\\
 & 6 & ReLu & $\medmuskip=0mu 4800\times1\times20$ & --\\
 & 7 & MPool (4,1), $s\!:$ 4 & $\medmuskip=0mu 1200\times1\times20$ & --\\ 
 & 8 & Conv (16,1) $\times$~20 & $\medmuskip=0mu 1200\times1\times20$ & $w\!:\!6400,b\!:\!20$\\
 & 9 & ReLu & $\medmuskip=0mu 1200\times1\times20$ & --\\
 & 10 & MPool (4,1), $s\!:$ 4 & $\medmuskip=0mu 300\times1\times20$ & --\\
 & 11 & Conv (8,1) $\times$~20 & $\medmuskip=0mu 300\times1\times20$ & $w\!:\!3200,b\!:\!20$\\
 & 12 & ReLu & $\medmuskip=0mu 300\times1\times20$ & --\\
 & 13 & APool (4,1), $s\!:$ 4 & $\medmuskip=0mu 75\times1\times20$ & --\\
 & 14 & MPool (5,1), $s\!:$ 5 & $\medmuskip=0mu 15\times1\times20$ & --\\
 & 15 & Global-Avg. & $\medmuskip=0mu 1\times1\times20$ & --\\
\midrule
\parbox[t]{2mm}{\multirow{4}{*}{\rotatebox[origin=c]{90}{Classifier}}} & 16 & FC 1, $n\!:$ 20 & $\medmuskip=0mu 1\times1\times20$ & $w\!:\!400,b\!:\!20$\\
 & 17 & FC 2, $n\!:$ 4 & $\medmuskip=0mu 1\times1\times4$ & $w\!:\!80,b\!:\!4$\\
 & 18 & Softmax  & $\medmuskip=0mu 1\times1\times4$ & --\\
 & ~-- & Class-output  & $\medmuskip=0mu 1\times1\times4$ & --\\
\bottomrule
\end{tabular}
\begin{flushleft} \footnotesize{keywords, Conv: Convolutional layer, ReLu: Rectified linear unit, MPool: Pooling by maximum operator, BNorm: Batch Normalisation, FC: Fully connected layer, APool: Pooling by averaging operator, $s$-stride, $w$-weights, $b$-bias, $o$-offset, $c$-scale, $n$-neuron} \end{flushleft}
\end{table}

The raw EEG signal from each channel is input to the proposed 1D CNN architecture. That is, the system is trained and tested  on each individual channel in the eight channel EEG. The EEG is divided into 5 minute segments with 50\% overlap. This 5-minute EEG segment was chosen by comparing the validation performance of epochs in the range of 0.5 to 10 minutes with a step size of 0.5 minutes.

Each 5-min EEG segment (19,200 samples) is input to a 1D convolutional layer. It evaluates a feature map by convolving the EEG segement with 10 different 1-second filters (64$\times$1); no padding is included in this convolution operation. The output of the convolutional layer is passed through the ReLu layer. Next, a max-pooling layer is included, which takes the maximum value from a sliding window segment (size 4$\times$1), with 4-stride downsampling. This results in a reduction in dimension from 19,200 to 4,800 sample-points across the 10-feature maps. The evaluated feature map is normalised across the 10-kernels by the batch normalisation layer.
The convolution, rectification, and pooling process is repeated another 3 times---see Table 1 for details. The final layer of this process is a global-average layer which produces only 1 sample point for all 20 feature maps. The classification then is performed by two fully connected layers each with 20 and 4 neurons respectively. Finally, the softmax layer provides the 1$\times$4 class probabilities. 

The integrated regularisation within the CNN provided by using shared weights and sparse connectivity results in fewer trainable parameters than fully connected networks, reducing the risk of overfitting. Hence, no early stopping criteria was found necessary in this study. The loss function used was categorical cross-entropy. Stochastic gradient descent was used with an initial learning rate of 0.01, this was reduced by 20\% every 5 iterations. Nesterov momentum was set to 0.9. A batch size of 128 was used for training and validation.

\subsection{Post-processing}
The CNN evaluates a probability vector for each processed 5-min EEG segment. 
A total of 23 EEG segments (based on 2.5 min overlap) are processed within 1-hour of EEG data. The total number of EEG segments varies based on the length of EEG data for each baby. We evaluate 2 different post-processing strategies to determine estimates of the HIE grade.

\subsubsection{One-step Voting}
\label{sssec:mepoch}
Class probabilities from 3 EEG segments are averaged over this 10-min period, resulting in 6 probability vectors per channel over a 1-hour epoch. This is repeated for all 8 channels. The overall probability vector is evaluated by averaging all $6 \times 8 =48$ probabilities from all 8-channels. 
The maximum global-average probability then determines the grade for the epoch. 

\subsubsection{Two-step Voting}
\label{sssec:lint}
First, the probability vectors from the averaged 10-minute segment in the one-step procedure is summarised across the 8 channels by taking the median value. This produces 6-successive 10-min segment vectors for a 1-hour epoch. Next, the probability vector is averaged over the 6 segments. The estimated grade is determined from the maximum global-average probability.


For training and testing, a leave-one-subject-out (LOSO) cross-validation method is used. Based on the 54 infants considered in study, system performance was observed by training the CNN model with 53 EEGs and testing on the one left out. This was repeated until all 54 infant's EEGs were tested to determine the overall classification accuracy of system.

\section{Results}
\label{expresult}

\begin{figure}[!h]
	\centering
	\includegraphics[width=0.98\linewidth]{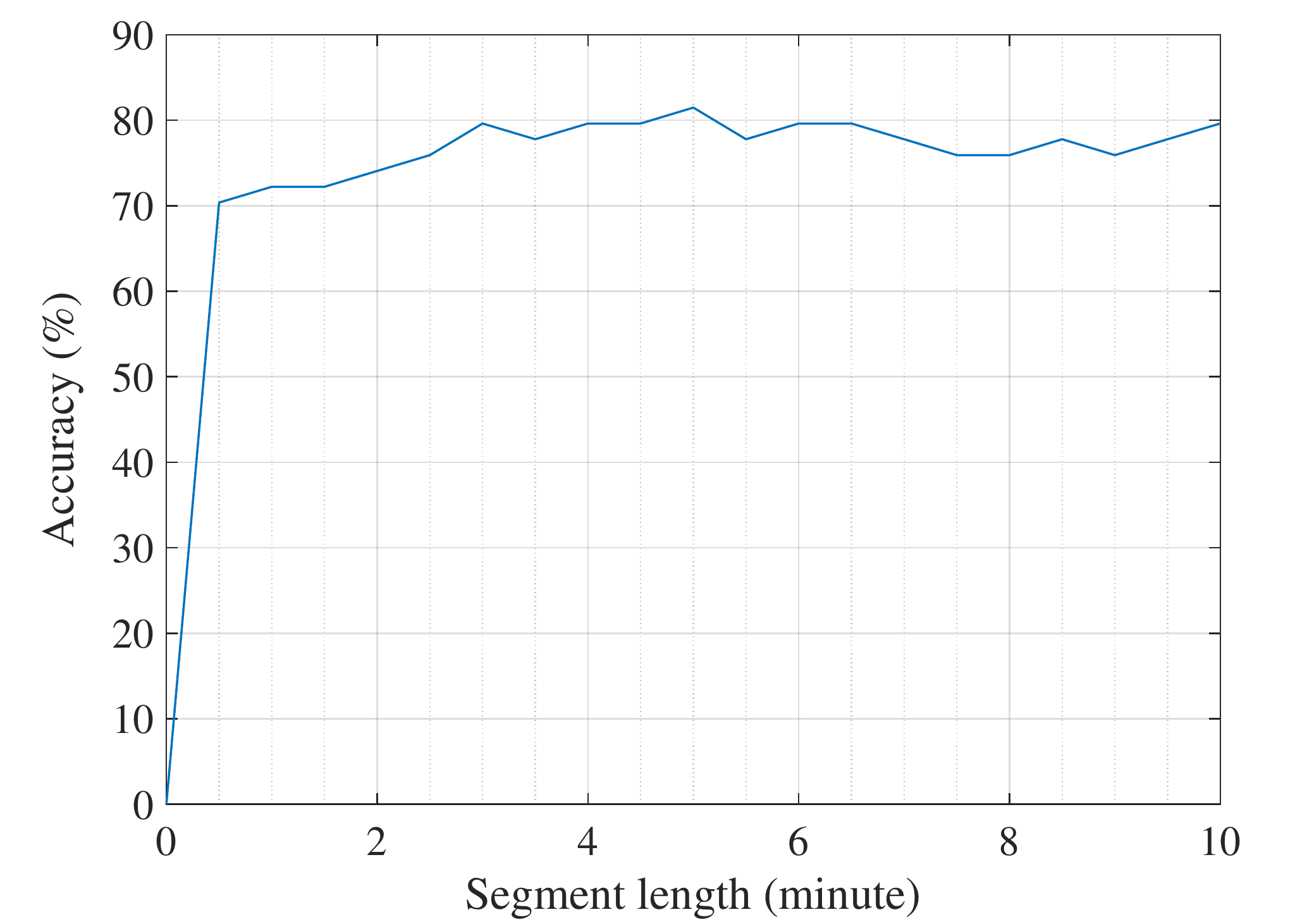}
	\caption{System performance using leave-one-out cross-validation with varying EEG segment length.}
	\label{sys_epoch}
\end{figure}

Fig. \ref{sys_epoch} illustrates system performance based on varying EEG segment length from 0.5 to 10 minutes with 50\% overlap. Performance is relatively stable for segments $\geq$2.5 minutes in duration, with accuracy $>$77\%. Highest classification accuracy, at 81.5\%, was achieved with a 5-minute segment.

\begin{figure}[h]
	\centering
	\includegraphics[width=1.0\linewidth]{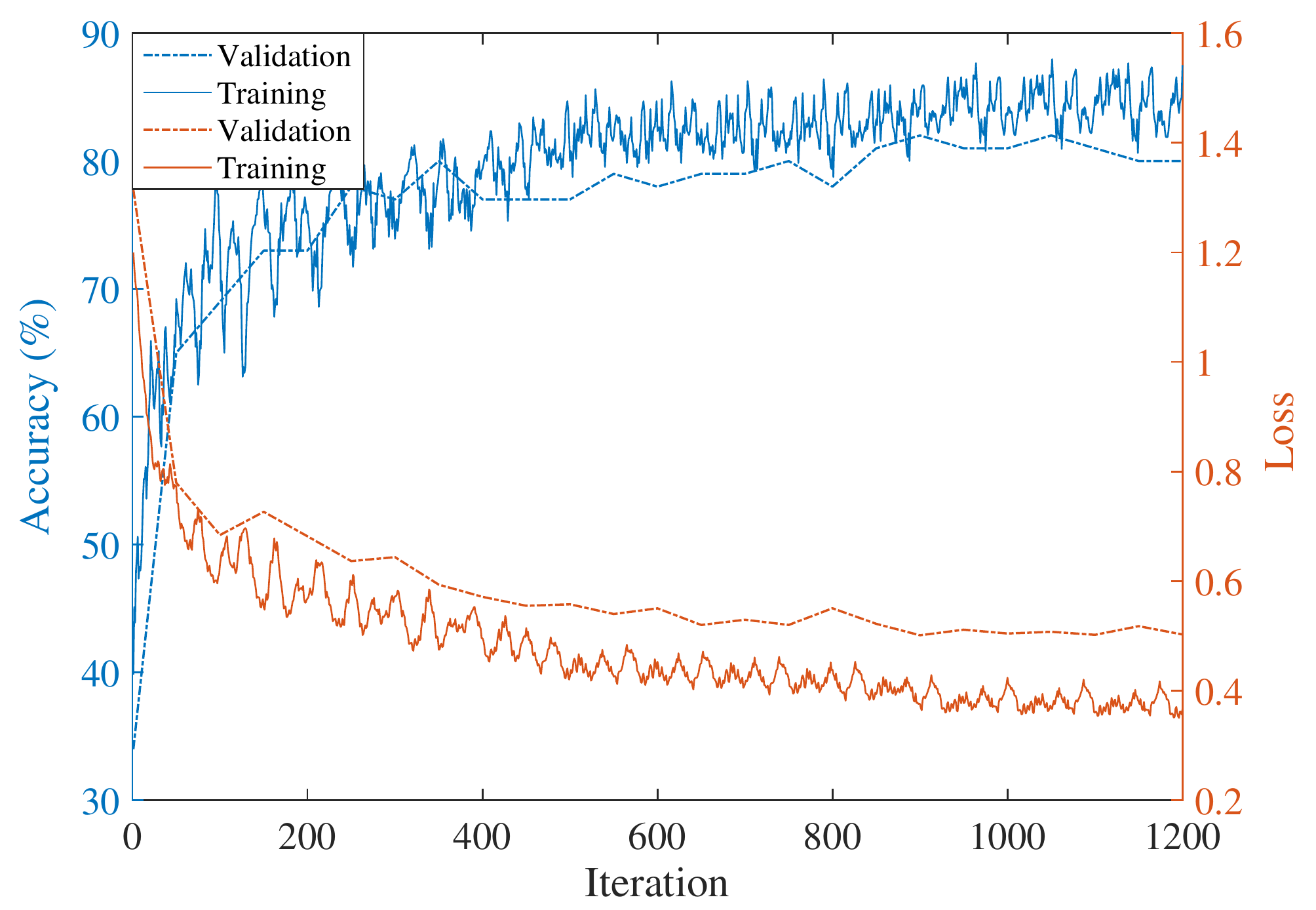}
	\caption{CNN accuracy and loss over training iterations for one iteration of a leave-one-out approach.}
	\label{train}
\end{figure}

Network training--validation performance over each training iteration is evaluated. It converges to a stable validation accuracy after approximately 1,000 learning iterations as shown in the training curve in Fig. \ref{train}.

The effect of the different post processing techniques on the system performance using the 5-min EEG segment is illustrated in Table \ref{postacc}. The CNN output is evaluated based on averaging the probability vector from total of 144 segments from all channels within an hour-long EEG. The one-step voting provides an overall decision based on 48 segments across eight channels on an hour-long EEG. Whereas the two-step voting considers the decision based on summarising across channels first and then across the 6 segments.

\begin{table}[!h]
\renewcommand{\arraystretch}{1.65}
\centering
\vspace{0.05cm}
\caption{Comparison of system performance based on different post processing techniques.}
\label{postacc}
\setlength\tabcolsep{12.0pt}
\begin{tabular}{r | c c c}
\toprule
\textbf{System} & CNN  & One-step & Two-step \\
\textbf{Performance} & Output & voting & voting\\
\midrule
\textbf{Accuracy} & 77.8\% & 79.6\% & 81.5\% \\ 
\bottomrule
\end{tabular}
\end{table}

The proposed two-step voting shows the highest accuracy for our data-set compared to the CNN output and one-step voting. The confusion matrix for the proposed system using two-step voting in LOSO cross-validation is presented in Table \ref{confusion}. It shows that 44/54 EEGs were correctly classified resulting in an accuracy of 81.5\%. In addition, no incorrect classifications were more than 1 grade away from the correct class.

\begin{table}[!h]
\renewcommand{\arraystretch}{1.40}
\centering
\caption{Confusion matrix based on the convolutional neural network output with two-step voting.}
\label{confusion}
\setlength\tabcolsep{2.85pt}
\begin{tabular}{c l l l l c r}
\midrule 
Actual& \multicolumn{4}{c}{System's Output} & \multirow{2}{*}{~Total~} & \multirow{2}{*}{~False}\\
 \cline{2-5}
grade & ~~~1 & ~2 & ~3 & ~4 & & \\ 
\midrule
1 & {\bf 22~(100\%)} & ~0 & ~0 & ~0 & 22 & 0~~~\\
2 & ~~~6 & {\bf 7~(50\%)} & ~1 & ~0 & 14 & 7~~~\\
3 & ~~~0 & ~3 & {\bf 9~(75\%)} & ~0 & 12 & 3~~~\\
4 & ~~~0 & ~0 & ~0 & {\bf 6~(100\%)} & 6 & 0~~~\\
\midrule
Total~ & ~~28 & 10 & 10 & ~6 & 54 & 10~~~\\
\midrule
\end{tabular}
\end{table}

The performance of existing, feature-based systems which use the same EEG dataset is compared with our proposed system; results are presented in Table \ref{compare}. The method from Stevenson \emph{et al}. \cite{R11} extracted a complex feature set from the instantaneous frequency and the instantaneous amplitude measures of a quadratic time-frequency distribution, combined using a linear classifier. Ahmed \emph{at al}. \cite{R12} uses a larger feature set of 55 features combined using super-vectors from a Gaussian mixture model in a support vector machine (SVM) classifier. Raurale \emph{at al}. \cite{R13} used two features of inter-burst activity evaluated from a burst detector and combined in a multi-layer perceptron. In contrast, our proposed system operates without the need for a feature extraction strategy and, although not as accurate as that developed by Ahmed \emph{et al.}, it still outperforms the Stevenson \emph{et al}. \cite{R11} and Raurale \emph{et al}. \cite{R13} systems.  

\begin{table}[h]
\renewcommand{\arraystretch}{1.5}
\centering
\caption{Comparison of the proposed method with existing techniques on the same database.}
\label{compare}
\setlength\tabcolsep{7.0pt}
\begin{tabular}{r | c c c c}
\toprule
\textbf{HIE Grading} & Stevenson & Ahmed & Raurale & Proposed \\
\textbf{Method} & \textit{et al}. \cite{R11} & \textit{et al}. \cite{R12} & \textit{et al}. \cite{R13} & method \\
\midrule
\textbf{Accuracy} & 77.8\% & 87.0\% & 77.8\% & 81.5\% \\ 
\bottomrule
\end{tabular}
\end{table}

\section{Discussion and Conclusions}
\label{discussion}
\vspace{0.10cm}
We present a novel deep-learning approach for grading HIE abnormalities in newborn EEG. We show how the CNN architecture can be applied to raw EEG to learn relevant features and classify the 4 HIE grades. The developed system has achieved comparable accuracy with some state-of-the-art HIE grading systems, without the need to develop complex feature sets. The proposed system with end-to-end optimisation of feature extraction and classification shows an advantage over traditional hand-crafted features based approaches. Future systems will incorporate more convolution layers in a deeper network for enhanced feature extraction with the aim of improving classification performance.

\end{document}